\documentclass[12pt]{cernlik}
\usepackage{epsfig}
\def\be{\begin{equation}}
\def\ee{\end{equation}}

\begin{document}
\begin{titlepage}
\title{TIME DISPERSION AND EFFICIENCY OF DETECTION
FOR SIGNALS IN GRAVITATIONAL WAVE EXPERIMENTS}

\author{P.Astone, S.D'Antonio and G. Pizzella\\
University of Rome Tor Vergata\\
INFN, Sezione di Roma\\
 and INFN, Laboratori Nazionali di Frascati, P.O. Box 13,
I-00044 Frascati, Italy
} 
\maketitle
\baselineskip=14pt
\begin{abstract}
Using simulated signals and measured noise with the EXPLORER and 
NAUTILUS detectors
we find the efficiency of signal detection and the signal
arrival time dispersion
versus the signal-to-noise ratio.
\end{abstract}
\vspace*{\stretch{2}}
\begin{flushleft}
	\vskip 2cm
{	PACS:04.80,04.30} 
\end{flushleft}
\end{titlepage}

\section{ Introduction}
 There are today five detectors of gravitational waves (GW) in operation 
\cite{expl,alle,niobe,naut,auri}, all of them of the resonant type.
It is thus important to study in detail the problem of the coincidence
search.

 In the past, after the initial works of Weber, three
papers on coincidence search have been published \cite{lrs,astro,lr}.
These coincidence search was made under two hidden assumptions:\\
a)the signal-to-noise ratio (SNR) was considered to be very large,\\
b)the event time was considered to be equal to the signal time.\\
Since we expect very tiny signals, the study of the problem
when dealing with small SNRs is fundamental.

This is our object here using simulated signals but with real
noise measured with the EXPLORER and NAUTILUS detectors.

\section{ Signal and events}
In order to clarify the distinction between $signal$ and $event$
let us recall how an $event$ is defined. We describe the procedure adopted
by the Rome group, but a similar procedure is adopted also by the
ALLEGRO, AURIGA and NIOBE groups.

For NAUTILUS and EXPLORER the
data have a sampling time of 4.544 ms and are filtered
with a filter matched to short bursts~\cite{fast}
for the detection of delta-like signals. The filter makes use of power
spectra obtained during periods of two hours.

Be $x(t)$ the filtered output of the electromechanical transducer
which converts the mechanical vibrations of the bar in electrical
signals. This quantity is normalized, using the detector calibration,
such that its square gives the energy innovation $E_f$
for each sample, expressed in kelvin units.
In absence of signals, for well behaved noise due only to the thermal
motion of the bar and to the electronic noise of the amplifier,
the distribution of $x(t)$ is normal with zero mean.
The variance (average value of the square of $x(t)$)
is called $effective~temperature$ and is indicated with $T_{eff}$. The
distribution of $x(t)$ is
\be
f(x)=\frac{1}{\sqrt{2\pi T_{eff}}}e^{\frac{x^2}{2T_{eff}}}
\label{normal}
\ee 

After the filtering of the raw-data, $events$ are extracted as follows.
A threshold is set in terms of a critical ratio defined by
\be
 CR=\frac{|x|-<|x|>}{\sigma(|x|)}=
\frac {\sqrt{SNR_f}-\sqrt {\frac {2}{\pi}}}{\sqrt {1-\frac {2}{\pi}}}
\label{creq}
\ee
where $\sigma(|x|)$ is the standard deviation of $|x|$ and
\be
 SNR_f=\frac{E_f}{T_{eff}}
\label{teffeq}
\ee

$T_{eff}$ is determined by
taking the average of the filtered data during the ten minutes preceeding
each considered event.
The threshold is set at CR=6, in order to have about one or two
hundred events per day. This corresponds to an energy
$E_t=19.5~ T_{eff}$. When the filtered data go above this threshold, the
time behaviour is considered until the filtered data
go below the threshold for
more than ten seconds. The maximum amplitude and its occurrence time
define the $event$.

By the word $signal$ here we mean the response of the detector to
an external excitation in absence of noise. It is then evident that
an $event$ is a combination of signal and noise. In the following
we shall use SNR to indicate the ratio between the $signal$ energy,
which we denote with $E_s$ and the noise $T_{eff}$,
\be
SNR=\frac{E_s}{T_{eff}}
\label{snr}
\ee
The effect of the noise on the signal has been discussed in
\cite{sta,alle,effi} and it turns out to be larger that one could
erroneously think. For example, with SNR=20 (for NAUTILUS), one
could think that most of the signals would be detected above
the threshold $E_t=19.5T_{eff}$. It turns out that the detection
efficiency is of the order of 50\%, as the noise might be in
phase with the signal, pushing it even higher over the threshold
or in counter-phase, pushing it below the threshold.
This means that the detection efficiency for  $m^{pl}$ 
coincidences with $m$ detectors,
in the case $E_s\sim E_f$, is of the order of $\frac{1}{2^m}$.

The noise acts also in producing an $event~ time$ different from
the time the $signal$ was applied. This influences
the choice of the coincidence time window.

\section{Experimental data}

We use two sets of experimental data, obtained with 
EXPLORER in 1991 and with NAUTILUS in 1998.

This is because the two detectors had their best performance
respectively in 1991 and 1998, and also because their
detection bandwidth is very different in the two cases.
The main characteristics of these two detectors are given in
table \ref{carac}.
\begin{table}
\centering
\caption{
Main characteristics of EXPLORER and NAUTILUS for the data used in
the present analysis.
}
\vskip 0.1 in
\begin{tabular}{|c|c|c|c|c|c|}
\hline
&year&temperature&Q&$\Delta f$&$T_{eff}$\\
\hline
EXPLORER&1991&2.6 K&$5~10^6$&1.9 Hz&6 mK\\
NAUTILUS&1998&0.15 K&$3~10^5$&0.12 Hz&4 mK\\
\hline
\end{tabular}
\label{carac}
\end{table}
The Q value for NAUTILUS is small because of electrical losses in
the transducer. Work is in progress to obtain a larger Q value.
Both EXPLORER and
NAUTILUS are equipped with similar resonant capacitive transducers, thus
they have two resonance modes at frequencies of 904.7 Hz and
921.3 Hz for EXPLORER and 907.0 Hz and 922.5 Hz for NAUTILUS.

\begin{figure}[hbt]
\vspace{12.0cm}
\centering
\includegraphics{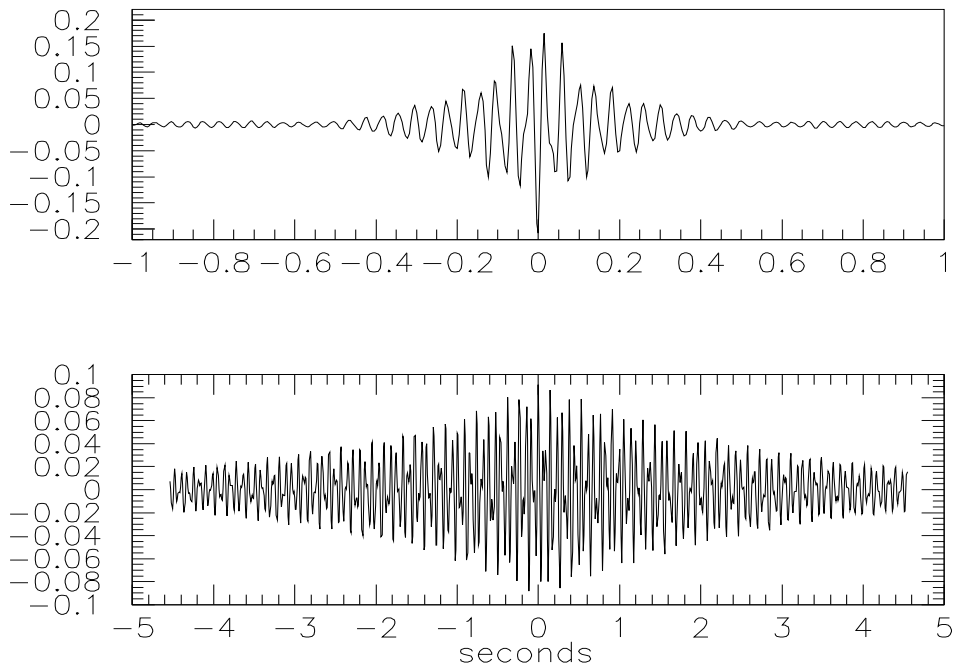}
 \caption{
Filtered data for a delta applied at the time 0.
Upper figure: EXPLORER. Lower figure: NAUTILUS. The decay time of the
envelope is measured to be 0.17 s for EXPLORER and 2.6 seconds
for NAUTILUS.
        \label{beta3t} }
\end{figure}

The algorithm for extracting small delta signals
from the noise is based on the measurement of the power spectra
and it takes care of both resonance modes \cite{fast}. Applying a delta
signal to the detector we have at the transducer output the
sum of the two mode oscillations, sharply beginning at the time
the pulse was applied and decaying with a time constant 
proportional to the Q value.
The filter operates a sort of weighted average and the
result V(t) has maximum value at the time the delta was applied
(t=0)
and oscillates, with envelope obeying the equation
\be
V(t)=V_o e^{-\beta_3|t|}
\label{beta3}
\ee
The quantity $\beta_3$ divided by $\pi$ gives the frequency bandwidth
of the apparatus.
An example of the behaviour of the filtered signal with time 
and in absence of noise is shown in fig.\ref{beta3t} for the two
detectors.

For the filtered data we get $\beta_3=6.0\frac{rad}{s}$ for EXPLORER
and $\beta_3=0.39\frac{rad}{s}$ for NAUTILUS. The bandwidth of EXPLORER in 1991
was then $\triangle f=\frac{\beta_3}{\pi}=1.9~Hz$, the bandwidth for
NAUTILUS in 1998 was $\triangle f=\frac{\beta_3}{\pi}=0.12~Hz$.
This small bandwidth will be increased in future with
improved transducers and electronics \cite{quantum}.

\section{Simulation with delta signals}

We make use of eight hours of data recorded with NAUTILUS on
12 July 1998 ($T_{eff}= 4.18~ mK)$
and we use four hours for EXPLORER recorded
on 13 September 1991 ($T_{eff}= 6.08~mK)$.

In absence of applied signals 34 events are detected
for EXPLORER and 41 events for NAUTILUS, due to
the noise fluctuation. These events are vetoed in all the successive
analyses made with applied signals.

Delta signals
with given $SNR$ are applied over the real noise with a certain
periodicity. One must make sure that the filtering of a new applied
signal is not disturbed by the residual of the previous applied
signal. This is obtained if the periodicity of the applied signals
is much larger than $\frac{1}{\beta_3}$. Thus we have used for
EXPLORER a periodicity
of half a minute for large SNR and a periodicity of ten seconds for
smaller SNR. For NAUTILUS the periodicities are one minute and
twenty seconds. The signals are applied at the exact time the data
are sampled with a sampling rate of 4.544 ms.

For EXPLORER we have found the result given in table \ref{effic1e}.
\begin{table}
\centering
\caption{
EXPLORER 1991.
Efficiency of detection, time deviation (one standard
deviation) and $\frac{E_f}{E_s}$ for 434 signals applied with periodicity of
half a minute and with various $SNR>5$. For SNR=2 and SNR=5 we have
applied 1300 signals with periodicity of 10 s (we have eliminated
from SNR=5 and 10 one event with time deviation of the order of
ten standard deviations).
}
\vskip 0.1 in
\begin{tabular}{|c|c|c|c|c|c|c|}
\hline
SNR&number of&detection&time&average of&theoretical\\
&detected&efficiency&deviation&$ES=\frac{E_f}{E_s}$&efficiency\\
&signals&\%&$[s]$&&\%\\
\hline
40&425&98&0.015&1.2&97.2\\
30&399&92&0.019&1.2&85.6\\
20&284&65&0.025&1.5&52.2\\
15&180&41&0.032&1.8&29.4\\
10&73 (74)&17&0.035&2.4&10.5\\
5&44 (45)&3.5&0.067&4.7&1.5\\
2&12&0.92&0.093&11&0.13\\
\hline
\end{tabular}
\label{effic1e}
\end{table}
For NAUTILUS we have found the result given in table \ref{effic1n}.
\begin{table}
\centering
\caption{
NAUTILUS 1998.
Efficiency of detection, time deviation (one standard
deviation) and $\frac{E_f}{E_s}$ for 448 signals applied with periodicity of
one minute and with various $SNR>5$. For SNR=2 and SNR=5 we have
applied 1328 signals with periodicity of 20 s.
}
\vskip 0.1 in
\begin{tabular}{|c|c|c|c|c|c|c|}
\hline
SNR&number of&detection&time&average of&theoretical\\
&detected&efficiency&deviation&$ES=\frac{E_f}{E_s}$&efficiency\\
&signals&\%&$[s]$&&\%\\
\hline
40&439&98&0.23&1.2&97.2\\
30&393&88&0.28&1.3&85.6\\
20&294&66&0.43&1.5&52.2\\
15&195&44&0.57&1.8&29.4\\
10&84&19&0.71&2.5&10.5\\
5&49&3.7&0.66&4.5&1.5\\
2&12&0.9&1.3&13&0.13\\
\hline
\end{tabular}
\label{effic1n}
\end{table}

\begin{figure}[b]
\vspace{12.0cm}
\centering
\includegraphics{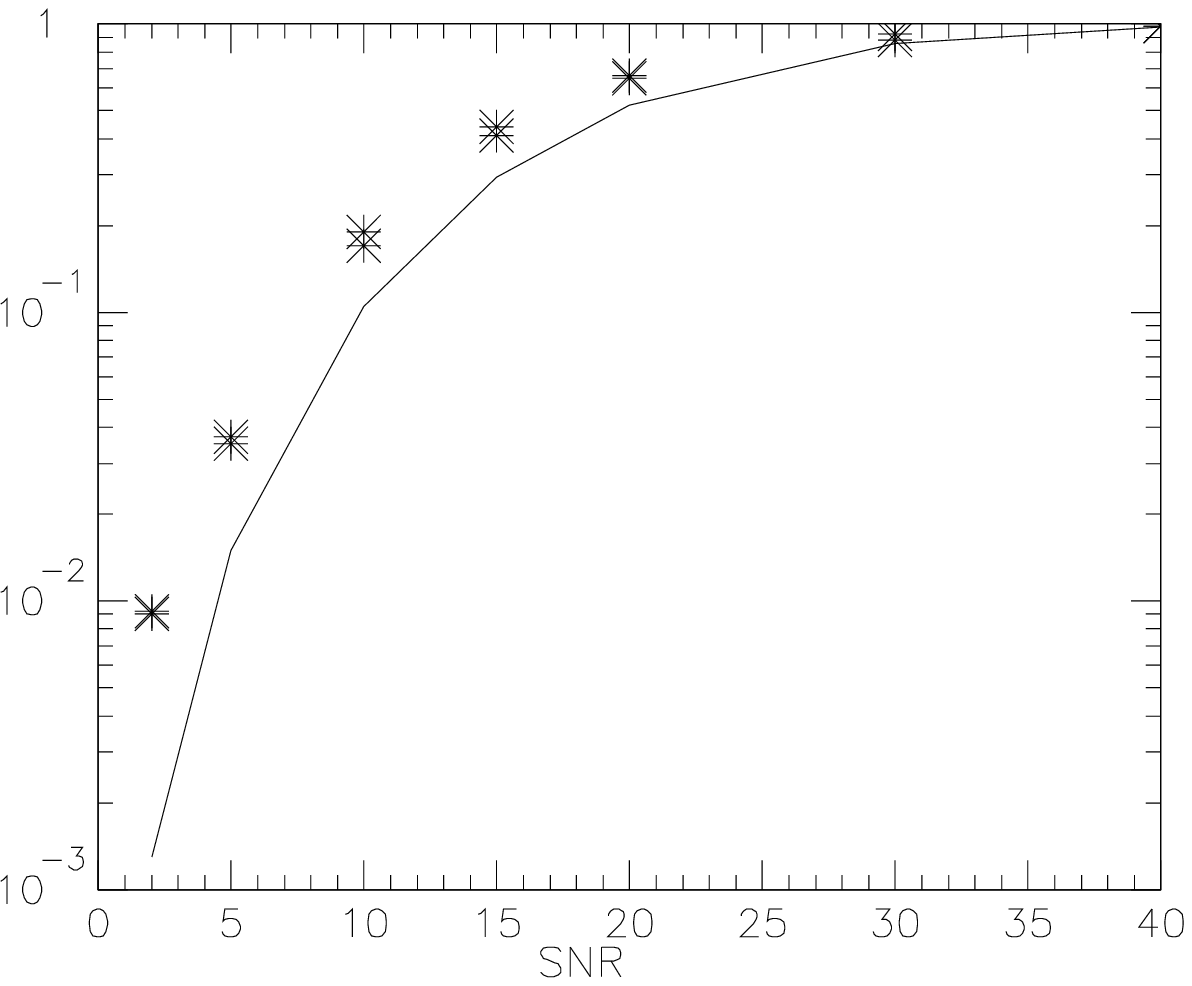}
 \caption{
The stars indicate the experimental efficiency for EXPLORER and
NAUTILUS versus SNR of the applied signals. The continuous line show
the expected theoretical efficiency as calculated with eq.\ref{papou}.
        \label{effi} }
\end{figure}

The efficiency is also shown in fig.\ref{effi}

The theoretical probability to detect a signal with a given SNR,
in presence of a well behaved Gaussian noise, is calculated as follows.
We put  $y=(s+x)^2$ where
$s\equiv \sqrt{SNR}$ is the signal we look for and $x$ is the gaussian
noise. We obtain easily \cite{papoulis}
\be
probability(SNR)=\int_{SNR_t}^{\infty} \frac {1}{\sqrt{2 \pi y}}
e^{-\frac{(SNR+y)}{2}}
cosh(\sqrt{y\cdot SNR})dy
\label{papou}
\ee
where we put $SNR_t=19.5$ for the present EXPLORER and NAUTILUS detectors.
     
The theoretical efficiency as deduced from eq.\ref{papou} is
reported in tables \ref{effic1e} and \ref{effic1n} and in fig.\ref{effi}.
 We notice a deviation between experimental
and theoretical efficiencies at small SNR. This is
due to the non gaussian character of the real noise.

\begin{figure}[b]
\vspace{12.0cm}
\centering
\includegraphics{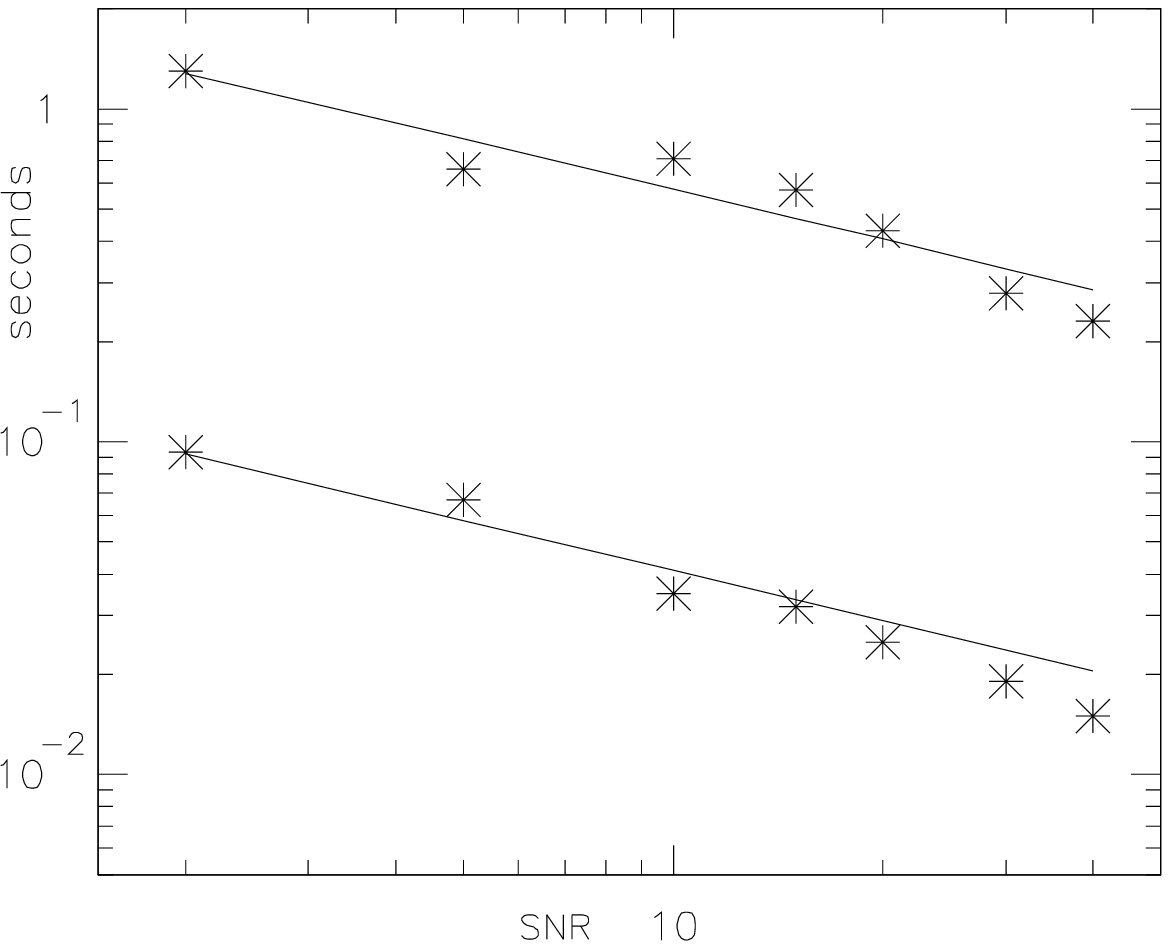}
 \caption{
Standard deviation of the event time with respect to the signal time,
versus SNR. The upper curve refers to NAUTILUS (bandwidth
$\Delta f=0.12~Hz$). The lower curve refers to EXPLORER
($\Delta f=1.9~Hz$). The lines are best fits with the eq. \ref{formula}.
        \label{deviation} }
\end{figure}

The time when the event due to a signal is observed deviates from
the time the signal is applied. We show in fig. \ref{deviation}
the standard deviation against SNR for EXPLORER 1991 and for NAUTILUS 1998.
The lines are the best fits with the following equations:\\
EXPLORER ~~~~~$\frac{1}{2\pi~1.74\pm 0.08}\sqrt{\frac{2}{SNR}}$\\
NAUTILUS ~~~~~$\frac{1}{2\pi~0.124\pm 0.007}\sqrt{\frac{2}{SNR}}$. 

We can write the empirical formula
\be
\sigma=\frac{1}{2\pi\Delta f}\sqrt{\frac{2}{SNR}}
\label{formula}
\ee

We see, as expected, that the time deviation decreases linearly with increasing
bandwidth. If we extrapolate to a SNR=100 and with a target
bandwidth for resonant detectors of the order of $\Delta f\sim 50~Hz$
we find a possible time resolution of the order of less than one
millisecond, as already recognized with room temperature
experiments \cite{vitale}.

\begin{figure}[b]
\vspace{12.0cm}
\centering
\includegraphics{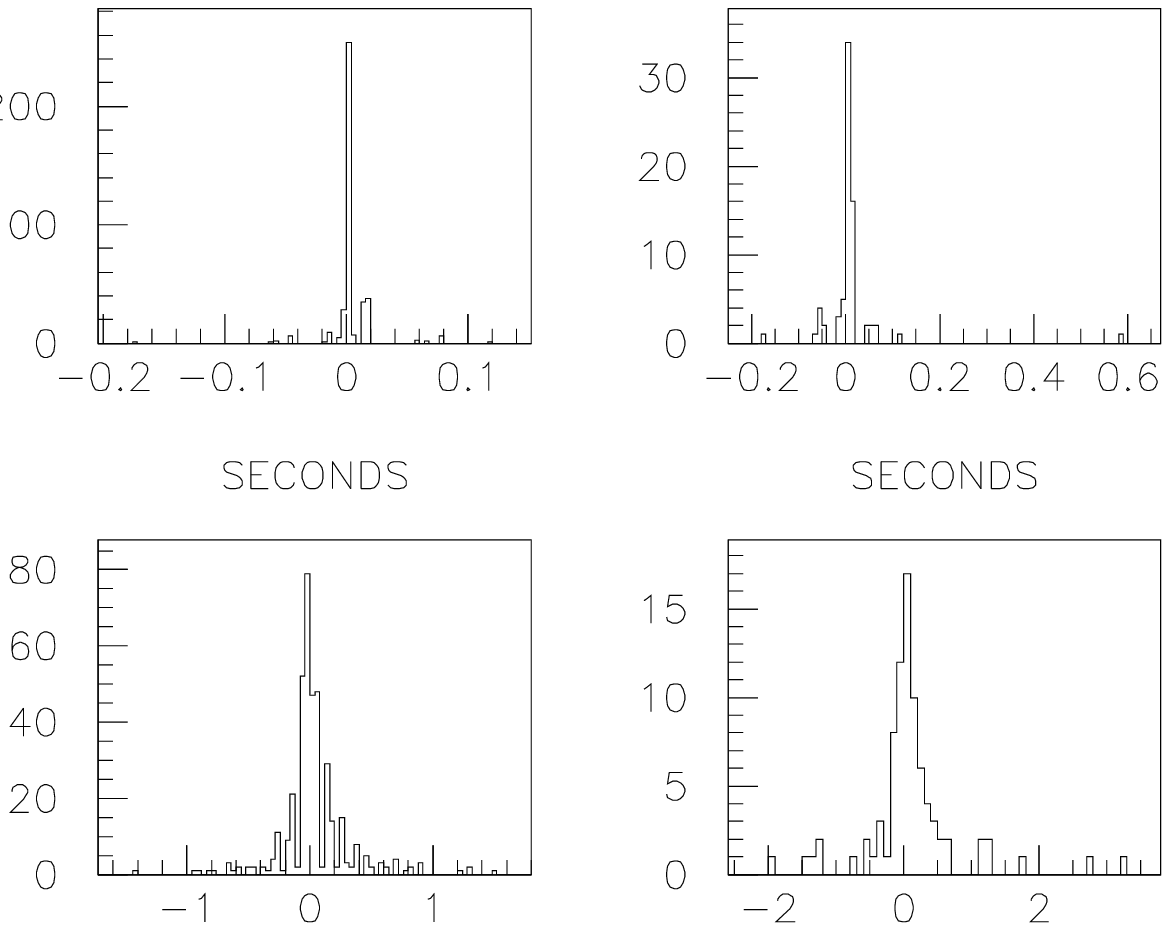}
 \caption{
Upper two figures for EXPLORER.
Delay distributions (time of the event minus the time the signal
was applied) for the detected delta signals with SNR=30 (left figure)
and SNR=10 (right figure).
Lower two figures for NAUTILUS.
        \label{distris30_10} }
\end{figure}

The delay distributions for signals with SNR=30 and SNR=10
are shown in fig. \ref{distris30_10}. We note for NAUTILUS 
a few events with delay greater than
1 s with respect to the time of the applied signals.
We have asked ourselves how it is possible to have a time deviation
over 1 s for signals with SNR=30. This is due to the fact that the noise,
although the data were selected so to have small 
$T_{eff}\leq5~mK$, has not completely a gaussian character.

\begin{figure}[t]
\vspace{12.0cm}
\centering
\includegraphics{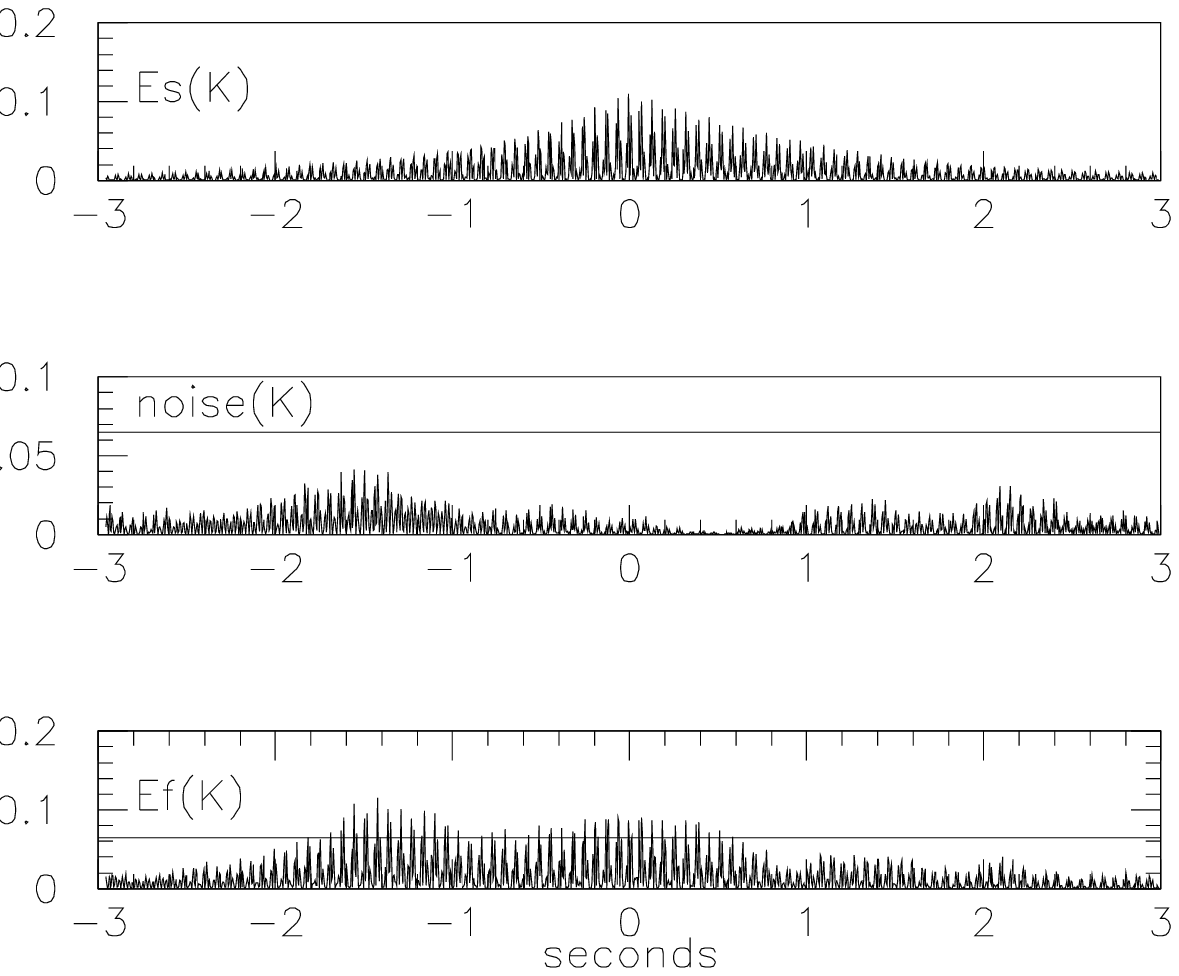}
 \caption{
NAUTILUS.
The upper figure shows the time behaviour of the applied signal energy
with SNR=30, in absence of noise, as in fig.\ref{beta3t}.
 The middle figure shows the noise time behaviour at the time this
signal was applied. The bottom figure shows the time behaviour
of the signal-plus-noise energy.
The horizontal line in the middle and lower figures indicates the
threshold energy $E_t=19.5~T_{eff}=65~mK$.
We see that, in this case, the maximum
value of the filtered data occurred 1.422 s before the delta signal
was applied (note that in the figure we give the energies,
but the signal and noise combine linearly with their amplitudes).
        \label{delta1s} }
\end{figure}

We have considered the particular case of the event
(fig.\ref{distris30_10}, SNR=30) detected 
with the NAUTILUS data 1.422 s $before$ the signal was applied.
In order to understand this result we plot
in fig.\ref{delta1s} the behaviour of the signal with
zero noise, of the noise alone and of the signal added to the noise.
For this particular case if we raise the signal to SNR=50 the corresponding
event has a time delay of -64 ms (still not quite zero).
If an additional filter is applied to the data, such as to require
i.e. that the detected event behaved in a gaussian way, the
signal with SNR=30 is lost, in spite of being a $delta$ signal.

\begin{figure}[hbt]
\vspace{12.0cm}
\centering
\includegraphics{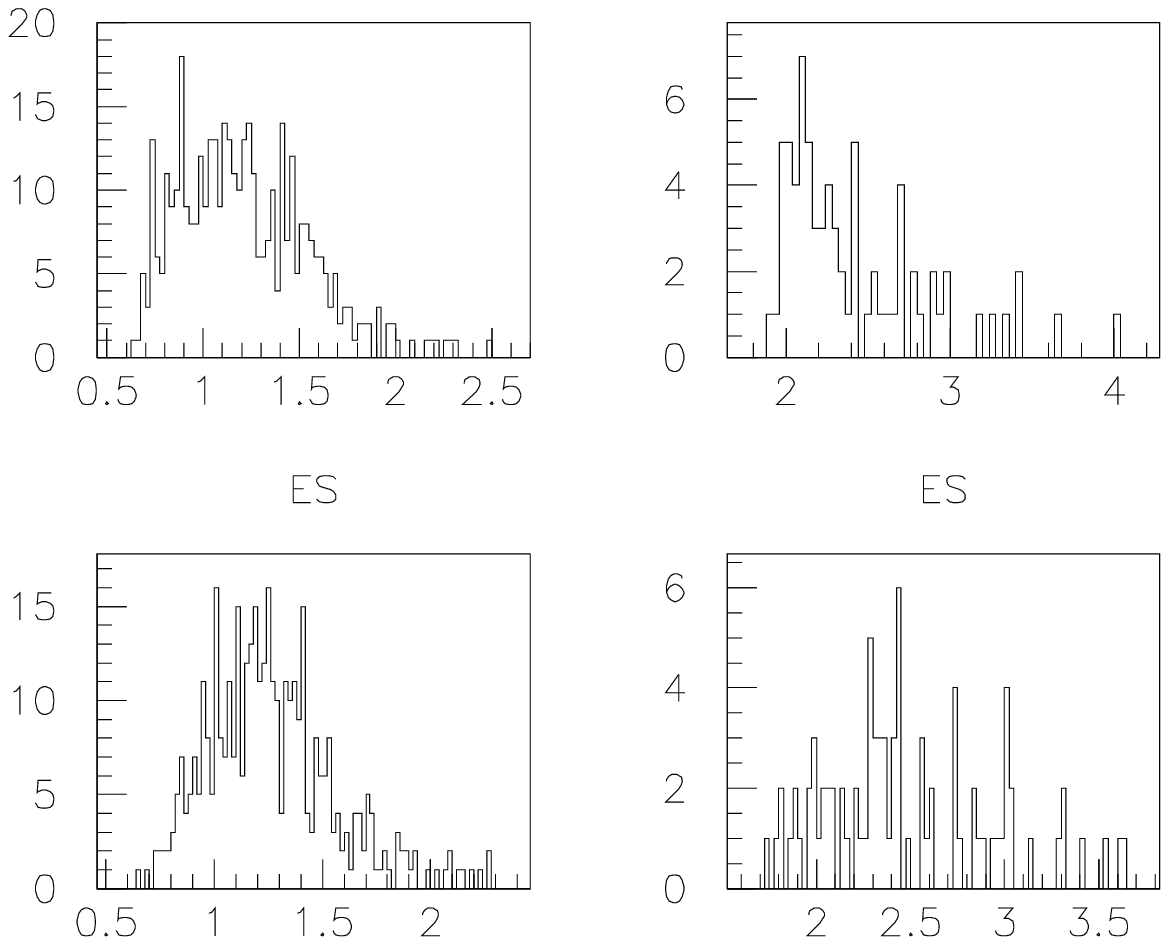}
 \caption{
The two upper figures refer to EXPLORER.
Distributions of $ES=\frac{E_f}{E_s}$ 
for the detected delta signals with SNR=30 (left figure)
and SNR=10 (right figure). $E_f$ is the energy of the event,
$E_s$ is the energy of the signal producing the event.
The two lower figures refer to NAUTILUS.
        \label{distrisnr} }
\end{figure}

We remark also that the events have energy
different from that of the signal. This is shown in 
tables \ref{effic1e},\ref{effic1n} and in fig.\ref{distrisnr}
where we give the distributions of the ratio $\frac{E_f}{E_s}$
for SNR=30 and SNR=10.

Finally we make the following consideration for the case when multiple
coincidences with $M$ detectors are searched for.
From tables \ref{effic1e} and \ref{effic1n}
 we deduce that when the signal is near the
threshold the efficiency of detection in nearly 50\%. This
means that for these signals the total efficiency
for $M^{pl}$ coincidences
is $\sim \frac{1}{2}^m$.  Since we want an efficiency
near unity (because of the very few possible GW signals)
we must consider only signals with $SNR$ at least twice the value $SNR_t$
of the threshold.

\section{ Conclusions} 

We have studied the $events$ generated in a resonant GW detector
when excited by GW bursts with $SNR$ near the threshold $SNR_t$ used for
defining the events. For $SNR=SNR_t$ the detection efficiency is nearly
$\frac{1}{2}$. The efficiency goes to 100\% for $SNR>2~SNR_t$, and it is
still $>10\%$ for $SNR\sim\frac{SNR_t}{2}$.

 The time of the event might be
different from that of the signal, with standard deviation 
depending on the SNR and on the bandwidth of the experimental
apparatus. In this analysis we have
applied delta signals at the exact time of the samples. If the delta
signals are applied randomly, as in the real case, the efficiency
will be smaller and the time dispersion larger.

 Delta-like signals can be lost if the requirement
to satisfy the theoretical behaviour expected for a delta signal
is imposed on the detected events, even for $SNR=30$,
as shown in fig.\ref{delta1s}. This can jeopardize a search looking
for very rare gravitational wave signals.

%
%

%
\end{document}